\documentclass[journal]{IEEEtran}

\usepackage{amsfonts}

\usepackage{ifpdf}
\usepackage{cite}
\ifCLASSINFOpdf
\usepackage[pdftex]{graphicx}
\else
\fi
\usepackage{amsmath}
\usepackage{algorithmic}
\usepackage{array}
\ifCLASSOPTIONcompsoc
  \usepackage[caption=false,font=normalsize,labelfont=sf,textfont=sf]{subfig}
\else
  \usepackage[caption=false,font=footnotesize]{subfig}
\usepackage{fixltx2e}
\usepackage{xcolor}
\usepackage{stfloats}

\usepackage{cite}
\usepackage{todonotes}
\usepackage{graphicx}
\usepackage{caption}

\newtheorem{Lemma}{Lemma}

\usepackage{float}
\floatstyle{ruled}
\newfloat{algorithm}{tbp}{loa}
\providecommand{\algorithmname}{Algorithm}
\floatname{algorithm}{\protect\algorithmname}
\usepackage{algorithmic}

\begin{document}

\title{\Huge Structural Identifiability of Impedance Spectroscopy Fractional-Order Equivalent Circuit Models\\With Two Constant Phase Elements}
\author{Tohid Soleymani Aghdam, Seyed Mohammad Mahdi Alavi, Mehrdad Saif 
\thanks{T. Soleymani Aghdam is with the Faculty of Electrical Engineering, Shahid Beheshti University, Daneshjou Boulevard, Tehran, Iran, Postal Code 1983969411. Email: {t\_soleymani@sbu.ac.ir}}
\thanks{S. M. M. Alavi is with the Department of Applied Computing and Engineering, School of Technologies, Cardiff Metropolitan University, Llandaff Campus, Western Avenue, Cardiff, UK, CF5 2YB. Email: {malavi@cardiffmet.ac.uk}}
\thanks{M. Saif is with the Department of Electrical Engineering, University of Windsor, Windsor, ON, Canada, N9B 3P4. Email: {msaif@uwindsor.ca}}
\thanks{M. Saif acknowledges the research funding from the Natural Sciences and Engineering Research Council of Canada (NSERC).}
\thanks{Matlab codes are available online at:\newline https://github.com/smmalavi/str-identifiability-frac-sys}
\thanks{{\em Corresponding Author}: S. M. M. Alavi }
}

\markboth{}%
{Shell \MakeLowercase{\textit{et al.}}: Bare Demo of IEEEtran.cls for IEEE Journals}

\maketitle

\begin{abstract}
Structural identifiability analysis of fractional-order equivalent circuit models (FO-ECMs), obtained through electrochemical impedance spectroscopy (EIS) is still a challenging problem. No peer-reviewed analytical or numerical proof does exist showing that whether impedance spectroscopy FO-ECMs are structurally identifiable or not, regardless of practical issues such as measurement noises and the selection of excitation signals. By using the coefficient mapping technique, this paper proposes novel computationally-efficient algebraic equations for the numerical structural identifiability analysis of a widely used FO-ECM with Gr\"{u}nwald-Letnikov fractional derivative approximation and two constant phase elements (CPEs) including the Warburg term. The proposed numerical structural identifiability analysis method is applied to an example from batteries, and the results are discussed. Matlab codes are available on github.
\end{abstract}


%
\IEEEpeerreviewmaketitle

\section{Introduction}
\label{sec:introduction}
Electrochemical impedance spectroscopy (EIS) is an important tool for the study of dynamics and properties of biomedical and electrochemical energy storage systems. EIS is used in \cite{Weiland2000,Mercanzini2009,Lempka2009,
Jiang2011,Venkatraman2011, Besancon2019} to study the {\em in vitro} and {\em in vivo} properties of electrode-tissue in neural stimulation techniques. In \cite{Halter2007} and \cite{Mishra2012}, EIS is used for differentiating between normal and malignant prostate tissues. 
Cancer detection by using EIS is studied in \cite{Kerner2002, Kao2008, Wang2012, Haeri2016, Han2016, Braun2017, Sevag-Packard2017, Zhang2018, Salahandish2018}. EIS has widely been used for monitoring of electrochemical energy storage systems in renewable energies and electric transportation. With EIS, electrochemical behaviors of batteries, super-capacitors and fuel-cells are estimated. It is shown that many specifications such as state of charge, age, internal temperature and resistance, and in general, state of health of electrochemical energy storage systems are identifiable by using EIS \cite{Buller2005, Blanke2005, Barsoukov2005, Troeltzsch2006, Richardson2014, deBeer2015, Zhu2015, Din2017, Guha2018, Howey2014, Alavi2015, Jacob2018, Tian2019, Ma2016, Xiong2018, Sabatier2015}.

In EIS, an excitation signal is applied to the system, and its response is measured. If excitation signal is current, the voltage response is collected or vice versa. The current and voltage signals are then employed for the computation of impedance spectra. The EIS data is further analyzed by fitting to an equivalent circuit model (ECM), \cite{Sadeghi2020} and references therein.

\begin{figure}[ht]
\centering
\includegraphics[scale=1]{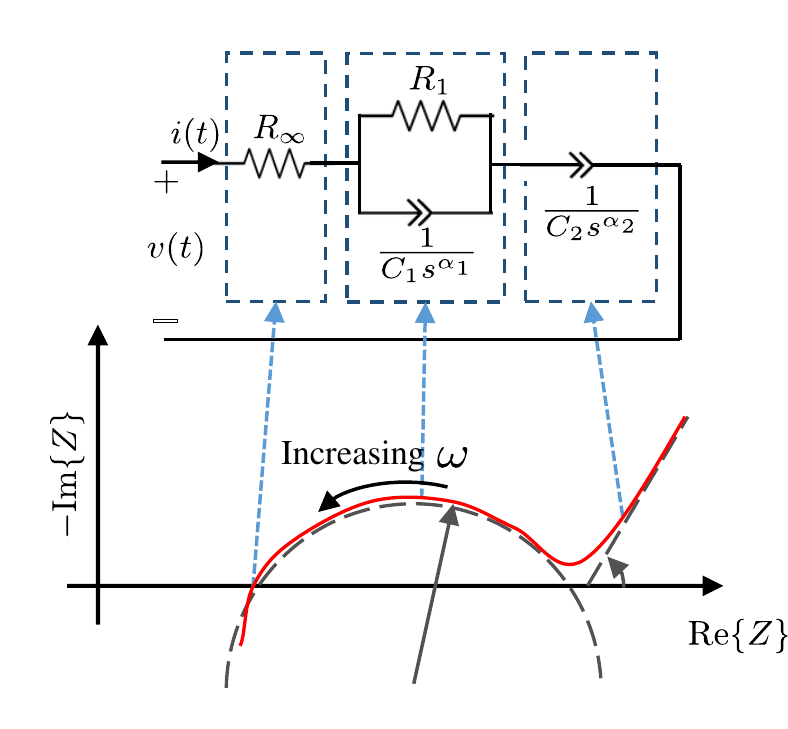}
\caption{The widely used fractional-order equivalent circuit model (FO-ECM) with two constant phase elements (CPEs) including the Warburg term, and corresponding impedance spectra.}
\label{fig:ecm2cpe}
\end{figure}

Figure \ref{fig:ecm2cpe} shows a typical EIS impedance spectra and its fractional-order ECM (FO-EM) that is the focus of this paper, and widely used in both biomedical and energy applications \cite{Weiland2000, Lempka2009, Jiang2011, Kao2008, Wang2012, Han2016, Sevag-Packard2017, Zhang2018, Salahandish2018, Barsoukov2005, Troeltzsch2006,
Buller2005, Guha2018, Jacob2018}. The intersection of the impedance spectra and the real axis represents the high-frequency ohmic resistance $R_{\infty}$. The mid-frequency semi-circle, modeled by a parallel $R_1$ and $1/(C_1s^{\alpha_1})$, and the low-frequency line, modeled by $1/(C_2s^{\alpha_2})$, represent charge transfer resistance, double layer capacitance, and  diffusion processes, where $s$ is the Laplace operator and $\alpha_i$'s, $i=1,2$,  are fractional exponents between 0 and 1, ($1 > \alpha> 0$). The term $1/{C_i s^{\alpha_i}}$ is referred to as the constant phase element (CPE), as its phase is constant with respect to frequency. The dimension of $C_i$, $i=1,2$ is $\mbox{Fcm}^{-2}s^{\alpha_i-1}$ \cite{Alavi2015}. The low frequency impedance  $1/(C_2s^{\alpha_2})$ is the so-called Warburg term. 

Estimation of the EIS FO-ECMs has been the topic of many papers. Usually, EIS impedance spectra is firstly computed and plotted in the Nyquist diagram by using Fourier transforms, and ECM parameters are then estimated by fitting to the impedance spectra in the frequency domain, \cite{Barsoukov2005}, \cite{Buller2005}, \cite{Howey2014}. An alternative approach estimates ECM parameters directly from the time-domain current and voltage signals by using system identification methods without Fourier transforms, \cite{Alavi2015, Alavi2017, Jacob2018, Sadeghi2020}. Estimation of the FO-ECM's parameters is beyond the scope of this paper, and interested readers are directed to \cite{Zou2018, Tian2020, Wang2019, Yang2017} for a more comprehensive survey of the proposed parameter estimation techniques.

The first essential stage in parameter estimation or system identification problems is to show that the model is structurally identifiable, regardless of practical issues such as measurement noises and the selection of excitation signals and parameter estimation method, \cite{Bates2019, Ljung1987, Soderstrom1989}. 

Since the 1970's, several analytical techniques have been proposed for structural identifiability analysis by using Taylor series expansion \cite{Pohjanpalo1978, Chappell1990}, similarity transformations \cite{Vajda1989, Anstett2008, Meshkat2014, Mahdi2014, Glover1974, DistefanoIII1977, VanDenHof1998}, differential algebra \cite{Ljung1994,Audoly2001}, and Laplace transforms (transfer functions) \cite{Cobelli1980, Bellman1970}. However, almost all of the proposed techniques deal with ordinary differential equations with integer orders.

This paper studies structural identifiability of the FO-ECM in Fig. \ref{fig:ecm2cpe}, with two CPEs under Gr\"{u}nwald-Letnikov derivative approximation, which is widely employed in the literature, \cite{Besancon2019, Guha2018, Jacob2018, Xiong2018,Ma2016, Tian2019}. In the arXiv preprint \cite{Alavi2015b}, the structural identifiability of the single-CPE FO-ECM is addressed by using the coefficient mapping technique. It is graphically proved that there is a one-to-one map between the coefficients of the transfer function and the parameters, and thus, the single-CPE FO-ECM is globally identifiable. It is further shown in \cite{Alavi2015b} that the structural identifiability of the two-CPE FO-ECM depends on the solution of the following set of equations:
\begin{equation}\label{2cpe-eq-alavi2015b}
\begin{aligned}
&g_1 + g_0(T+1)\left(\frac{1}{\alpha_1-T}+\frac{1}{\alpha_2-T}\right) = 0\\
&g_2  - g_0(T+1)(\hat{a}+\hat{b}+\hat{c}) = 0,
\end{aligned}
\end{equation}
where $g_i$'s, $i=0,1,2$ are the lowest orders' coefficients of the ECM transfer function, $T$ is the total number of samples taken from the voltage and current signals, and
\begin{align*}
&\hat{a}=\frac{T}{(\alpha_2-T)(\alpha_2-T+1)},~\hat{b}=\frac{(T+1)}{(\alpha_1-T)(\alpha_2-T)}\\
&\hat{c}=\frac{T}{(\alpha_1-T)(\alpha_1-T+1)}.
\end{align*}
However, these equations suffer from an ill-conditioned problem that is: for large $T$'s, $g_i$'s tend to zero. Under the ill-conditioned problem, it is hard to solve \eqref{2cpe-eq-alavi2015b}, and deduce the structural identifiability.

\subsection{The main contribution of this paper}
The main contribution of this paper is to propose alternative set of equations which do not face with the ill-conditioned problem. The new equations are computationally more efficient, compared to those in \cite{Alavi2015b}, and their numerical solution determines whether the two-CPE FO-CEM is structurally identifiable or not. The proposed numerical structural identifiability analysis method is applied to the battery EIS FO-ECMs and the results are discussed.

\subsection{Assumptions}
The following assumptions are made throughout this paper.
\begin{itemize}
\item[-] Due to complexities, this paper only addresses the two-CPE FO-ECMs with Gr\"{u}nwald-Letnikov fractional derivative approximation. Structural identifiability analysis for higher-order FO-ECMs and/or by using other approximations of fractional derivatives remains an open problem.
\item[-]This paper only focuses on the structural identifiability analysis, dealing with the model structure under the noise-free condition. The practical identifiability analysis, i.e., the selection of the excitation signals and parameter estimation technique, and the effect of measurement noise are beyond the scope of this paper. In \cite{Nazarian2010}, practical identifiability of fractional commensurate-order models is studied, where all differential orders are integer multiples of a base order. The proposed analysis is in the frequency domain, showing that fractional commensurate-order models are poorly identifiable for small values of the base order, \cite{Nazarian2010}. In \cite{Jacob2018}, practical identifiability of the single-CPE and two-CPE FO-ECMs are studied, and the results confirm that the input signal plays a key role in the parameter estimation. The selection of the excitation signal for the identification of FO-ECMs is still an open problem. In \cite{Sharma2014, Rothenberger2014, Alavi2017, Sadeghi2020}, a number of methods are proposed for the selection of the excitation signal for the identification of the ordinary-order ECMs.
\end{itemize}

\subsection{The structure of the paper}
This paper is organized as follow. Section II describes FO-ECMs using the Gr\"{u}nwald-Letnikov approximation. In section III, the proposed structural identifiability analysis method is described. Numerical results are given and studied in section IV for an example from battery applications.

\section{Model Structure}\label{sec:model}
By using the Gr\"{u}nwald-Letnikov approximation, a transfer function of the FO-ECM in Fig.  \ref{fig:ecm2cpe} is given by, \cite{Alavi2015b, Jacob2018}:
\begin{align}
\label{exRRCC:tf1}
H(z,\theta)=d(\theta)+\displaystyle \sum_{i=1}^{2}\frac{b_i(\theta)z^{T}}{z^{T+1}- \sum_{j=0}^{T} a_{i,j}(\theta)z^{T-j}}
\end{align}
where, $z$ is the discrete-time operator, $i$ denotes the $i-$th CPE, $j$ denotes the $j-$th sample, $T$ is the total number of samples, and 

\begin{eqnarray}\label{gen-ecm-params}
\begin{aligned}
& \theta=\{R_\infty,~R_1,~C_1,~\alpha_1,~C_2,~\alpha_2\}, ~d(\theta)=R_\infty,\\&b_i(\theta)=\frac{T_s^{\alpha_i}}{C_i},~ a_{1,0}(\theta)= \alpha_1-\frac{T_s^{\alpha_1}}{R_1C_1},\\& a_{2,0}(\theta)= \alpha_2,~a_{i,j}(\theta)= -(-1)^{j+1}\binom{\alpha_i}{j+1},\\
& i=1,2,~j=1,2,\cdots,T,
\end{aligned}
\end{eqnarray}
where, $\binom{\alpha_i}{j}$ is the binomial coefficient given by
\begin{equation*}\label{Gamma_func}
\binom{\alpha_i}{j}=\frac{\Gamma(\alpha_i+1)}{\Gamma(j+1)\Gamma(\alpha_i+1-j)},
\end{equation*}
and, $\Gamma(\cdot)$ denotes the gamma function.

By the expansion of \eqref{exRRCC:tf1}, a monic transfer function, where the coefficient of the highest order term in the denominator is 1, is given by, \cite{Alavi2015b, Jacob2018}:
\begin{align}
\label{exRRCC:tf2}
&H(z,\theta)= \frac{f_{2T+2}(\theta)z^{2T+2}+f_{2T+1}(\theta)z^{2T+1}+\cdots+ f_{0}(\theta)}{z^{2T+2}+g_{2T+1}(\theta)z^{2T+1}+\cdots+ g_{0}(\theta)}.
\end{align}

\section{Structural Identifiability Using Coefficient Mapping}\label{sec:strident}
The structural identifiability analysis, by using the coefficient mapping method, is based on the following lemma.

\begin{Lemma}[\cite{Mahdi2014}]\label{lem:Coefmap}
Consider a model structure $\mathcal{M}(\theta)$, where $\theta$ represents the parameter vector. Let assume that a transfer function of the model structure is given by \eqref{exRRCC:tf2}. Then, the model structure $\mathcal{M}(\theta)$ is
\begin{itemize}
\item[-]  \emph{globally identifiable} if there is a one-to-one map between the coefficients of the transfer function and the parameter vector $\theta$,
\item[-] \emph{identifiable} if there is a many-to-one map between the coefficient of the transfer function and the parameter vector $\theta$,
\item[-] \emph{unidentifiable} if there is an infinitely many-to-one map between the coefficient of the transfer function and the parameter vector $\theta$.\hfill{$\square$}
\end{itemize}

\end{Lemma}

The one-to-one map between the coefficients of the transfer function and the parameter vector $\theta$ means that the parameters $R_\infty,~R_1,~C_1,~\alpha_1,~C_2,$ and $\alpha_2$ are uniquely computed given the coefficients of the transfer function. Likewise, the (infinitely) many-to-one map  between the coefficients of the transfer function and the parameter vector $\theta$ means that coefficients of the transfer function result in (infinitely) many sets of $R_\infty,~R_1,~C_1,~\alpha_1,~C_2,$ and $\alpha_2$.


In order to overcome the computational issue seen in \cite{Alavi2015b}, the coefficients with highest indices, i.e., $f_{2T+2}$, $f_{2T+1}$, $g_{2T+1}$, $f_{2T}$, $g_{2T}$, etc., are processed in this paper. The denominator coefficients are given by:
\begin{align}
g_{2T+1}(\theta)&=-\left(a_{1,0}+a_{2,0}\right) \label{eq3} \\
g_{2T}(\theta)&=-\left(a_{1,1}+a_{2,1}\right)+a_{1,0}a_{2,0} \label{eq4} \\
g_{2T-1}(\theta)&=-\left(a_{1,2}+a_{2,2}\right)+a_{1,0}a_{2,1}+a_{1,1}a_{2,0} \label{eq5} \end{align}
From \eqref{gen-ecm-params},
\begin{align}
\label{a20}
a_{2,0}=\alpha_2
\end{align}
By using \eqref{eq3} and \eqref{a20}, it is deduced that:
\begin{align}
a_{1,0}=-\left(g_{2T+1}+\alpha_2\right) \label{eq7}
\end{align}
From Lemma 3 in \cite{Alavi2015b}, the relationship between $a_{i,j}$ and $a_{i,j+1}$ is given by:
\begin{align}
a_{i,j+1}=-\frac{\alpha_i-j-1}{j+2}a_{i,j} \,\,\,\mbox{for}~j\geq1
\label{eq6}
\end{align}
By using \eqref{a20}, \eqref{eq7}, and \eqref{eq6}, the manipulation of $g_{2T}$ and $g_{2T-1}$ results in:
\begin{align}
a_{1,1}&=\frac{B}{C}\\
a_{2,1}&=A-\frac{B}{C}
\end{align}
where,
\begin{align}
 \nonumber B&=g_{2T-1}-\Big(\alpha_2(g_{2T+1}+\alpha2)+g_{2T}\Big)\Big(g_{2T+1}+\displaystyle\frac{2\alpha_2+2}{3}\Big)\\
C&= g_{2T+1}+\displaystyle\frac{\alpha_1+5\alpha_2}{3}\\
A&=-\Big(\alpha_2(g_{2T+1}+\alpha2)+g_{2T}\Big)
\end{align}
Thus, $a_{1,0}$, $a_{2,0}$, $a_{1,1}$, and $a_{2,1}$ are written in terms of $\alpha_1$ and $\alpha_2$.
From the coefficients of the numerator, it is deduced that:
\begin{align}
f_{2T+1}-dg_{2T+1}&=b_1+b_2 \label{eq10} \\
f_{2T}-dg_{2T}&=-b_1a_{2,0}-b_2a_{1,0} \label{eq11} \end{align}
The solutions of \eqref{eq10} and \eqref{eq11} are given by:
\begin{align}
b_1&=\frac{D}{E},\mbox{~and}
\label{eq12}\\
b_2&=(f_{2T+1}-dg_{2T+1})-b_1 \label{eq13} \end{align}
where,
\begin{align*}
D&=(f_{2T+1}-dg_{2T+1})(g_{2T+1}+\alpha_2)-(f_{2T}-dg_{2T})\\
E&=g_{2T+1}+2\alpha_2
\end{align*}
For the next two coefficients, i.e., $f_{2T-1}$ and $f_{2T-2}$, the following equations hold:
\begin{align}
f_{2T-1}-dg_{2T-1}=-b_1a_{2,1}-b_2a_{1,1}\label{eq14}\\
f_{2T-2}-dg_{2T-2}=-b_1a_{2,2}-b_2a_{1,2}\label{eq15}
\end{align}

The replacement of $b_1$, $b_2$, $a_{1,0}$, $a_{2,0}$, $a_{1,1}$ and $a_{2,1}$ in \eqref{eq14} and \eqref{eq15} yields:
\begin{align}
\nonumber & (f_{2T+1}-dg_{2T+1})BE+(f_{2T-1}-dg_{2T-1})CE+\\&\hspace{10em}ACD-2BD =0 \label{eq16}\\ \nonumber
&(f_{2T+1}-dg_{2T+1})(\alpha_1-1)BE+(\alpha_2-2)ADC-\\&(\alpha_1+\alpha_2-4)BD-3(f_{2T-2}-dg_{2T-2})CE=0 \label{eq17} \end{align}

In \eqref{eq16} and \eqref{eq17}, all terms of $A$, $B$, $D$ and $E$ are only dependent on $\alpha_2$. Thus, $\alpha_1$ is expressed as a function of $\alpha_2$ by solving and rearranging \eqref{eq16} and \eqref{eq17}, which lead to:
\begin{equation}
\alpha_1=\frac{K_1^4(\alpha_2)}{K_2^3(\alpha_2)}=\frac{K_3^5(\alpha_2)}{K_4^4(\alpha_2)}\label{eq18}\end{equation}
where $K_i^j$ represents the $i$th polynomial with degree of $j$.

The exponent $\alpha_2$ is found by solving the following equation.
\begin{align}
\frac{K_1^4(\alpha_2)}{K_2^3(\alpha_2)}-\frac{K_3^5(\alpha_2)}{K_4^4(\alpha_2)}=0\label{eq19}
\end{align}
Since the resulting equation is of order eight, it is only possible to solve it numerically. If there is at least one set of $\alpha_i \in (0~1)$, $i=1,2$, then the model is identifiable. After obtaining $\alpha_i \in (0~1)$, $i=1,2$, other parameters are found by calculating $a_{1,0}$, $b_1$ and $b_2$ and equations in \eqref{gen-ecm-params}.

The parameters $b_1$ and $b_2$ are positive. The $\alpha_2$ value between $(f_{2T}-dg_{2T})/(f_{2T+1}-dg_{2T+1})-g_{2T+1}$ and $-(f_{2T}-dg_{2T})/(f_{2T+1}-dg_{2T+1})$ leads to negative  $b_1$ and $b_2$. Thus, acceptable $\alpha_2$'s meet the following creterion.
\begin{align}
\alpha_2  \notin \Big(\frac{f_{2T}-dg_{2T}}{f_{2T+1}-dg_{2T+1}}-g_{2T+1},-\frac{f_{2T}-dg_{2T}}{f_{2T+1}-dg_{2T+1}}\Big)\label{eq20}
\end{align}


If more coefficients of the transfer function are considered, equations with higher degrees are achieved. However, their solutions must meet the equation \eqref{eq19}. Thus, \eqref{eq19} is the lowest order equation that is required for structural identifiability analysis.

\section{Results}
The proposed structural identifiability analysis method is applied to a battery cell FO-ECM in Fig.  \ref{fig:ecm2cpe} with $\alpha_1=0.8,~\alpha_2=0.5, ~R_\infty=0.01,~ R_1=0.2,~ C_1=3,$ and $C_2=400$. These values are within the range commonly used in litretaure, \cite{Jacob2018}. By using the equations in \eqref{gen-ecm-params}, the coefficients of the original transfer function are computed and given in Table \ref{tab2}. The question is that whether it is possible to compute $\alpha_1,~\alpha_2, ~R_\infty,~ R_1,~ C_1,$ and $C_2$, given the coefficients of the original transfer function? It is recalled that structural identifiability is a noise-free concept.
\\
\begin{table}
\caption{transfer function coefficient}
\label{table}
\centering
\setlength{\tabcolsep}{3pt}
\begin{tabular}{|p{45pt}|p{60pt}|p{45pt}|p{45pt}|}
\hline
Num. Coef.& Value& Den. Coef.& Value\\ \hline
$f_{2T+2}$ & $0.01$ & $g_{2T+1}$ & $-1.2962$\\ \hline
$f_{2T+1}$& $-0.0121$ & $g_{2T}$& $0.1931$\\ \hline
$f_{2T}$& $0.0015$ & $g_{2T-1}$& $0.0450$\\ \hline
$f_{2T-1}$& $3.505\times 10^{-4}$ & $g_{2T-2}$& $0.0191$\\ \hline
$f_{2T-2}$& $1.416\times 10^{-4}$ & $g_{2T-3}$& $0.0103$\\ \hline
$f_{2T-3}$& $7.218\times 10^{-5}$ & $g_{2T-4}$& $0.0063$\\ \hline
\end{tabular}
\label{tab2}
\end{table}

In order to increase the accuracy of computations, the Matlab command `vpa(x,digits)' is used, where  `digits' denotes the digit accuracy.

At the first step, solutions of the exponent $\alpha_2$ are computed by solving the following equation that is obtained from \eqref{eq19}:
\begin{multline*}
\alpha_2^8-5.395708923047713\alpha_2^7\\
+12.451808248913298\alpha_2^6-16.088049799882121\alpha_2^5\\
+12.743527275051907\alpha_2^4-6.338984994985100\alpha_2^3\\
+1.932660443044634\alpha_2^2-0.329710967652997\alpha_2\\
+0.024032821066090=0
\end{multline*}

Two complex roots are obtained, which are not acceptable. The real roots are listed in Table \ref{tab3}.

\begin{table}
\caption{Real Solutions for $\alpha_1$ and $\alpha_2$}
\label{table}
\centering
\setlength{\tabcolsep}{3pt}
\begin{tabular}{|p{35pt}|p{87pt}|p{87pt}|}
\hline
Roots& $\alpha_2$ & $\alpha_1$\\ \hline
pair 1& $0.298245954619025$& $2.397337600606689$\\ \hline
pair 2& $0.500000000000000$ & $0.800000000000000$ \\ \hline
pair 3& $0.625975537273579$& $0.677356198694181$ \\ \hline
pair 4& $0.646678864697306$& $0.655173050215288$ \\ \hline
pair 5& $0.797894050107465$& $0.499243173767398$ \\ \hline
pair 6& $1.295547992101849$ & $-2.589172586806396$ \\ \hline
\end{tabular}
\label{tab3}
\end{table}
Among the pairs listed in Table \ref{tab3}, four pairs are located in interval $(0,1)$, which are acceptable. From \eqref{eq20}, $\alpha_2$ in range of (0.52024,0.77595) leads to non-positive value for $b_{1}$ and $b_2$, and can thus be removed. Only two pairs, pair2 and pair5 in Table \ref{tab3}, are left at this stage. The analysis is continued by comparing the transfer function built from the remaining two pairs and the original transfer function. Table \ref{tab4} shows the normalized errors between the coefficients of the estimated and original transfer functions, $\Delta g_{i}$, for both pair2 and pair5, where $i$ denotes the index of the coefficient. It is seen that the solution pair2 is only acceptable, because of negligible error achieved. Only one answer was then found for both $\alpha_1$ and $\alpha_2$, thus the model is globally structurally identifiable. Other system parameters are easily calculated after that $\alpha_1$ and $\alpha_2$ are found.

\begin{table*}
\caption{ Normalized errors between the coefficients of the estimated and original transfer functions, for both pair2 and pair5.}
\label{table}
\centering
\setlength{\tabcolsep}{3pt}
\begin{tabular}{|p{30pt}|p{55pt}|p{55pt}|p{55pt}|p{55pt}|p{55pt}|p{55pt}|p{55pt}|}
\hline
& $\Delta g_{2T} $ & $\Delta g_{2T-1} $ & $\Delta g_{2T-2} $ & $\Delta g_{2T-3}$ & $\Delta g_{2T-4} $ & $\Delta g_{2T-5} $ & $\max\{\Delta g_{i}\}$ for $i=1:15$ \\ \hline
pair 2 & $4.7559e^{-40}
$ & $1.01986e^{-39}$ & $1.20208e^{-39}$ & $2.23045e^{-40}$ & $9.06007e^{-40}$ & $3.42014e^{-40}$ & $2.63764e^{-39}$ \\ \hline
pair 5 & $0.00598$ & $0.00112$ & $0.00779$ & $0.01049$ & $0.012981$ & $0.030246$ & $0.03169$ \\ \hline
\end{tabular}
\label{tab4}
\end{table*}

\begin{figure}[h]
\centering
\includegraphics[scale=.55]{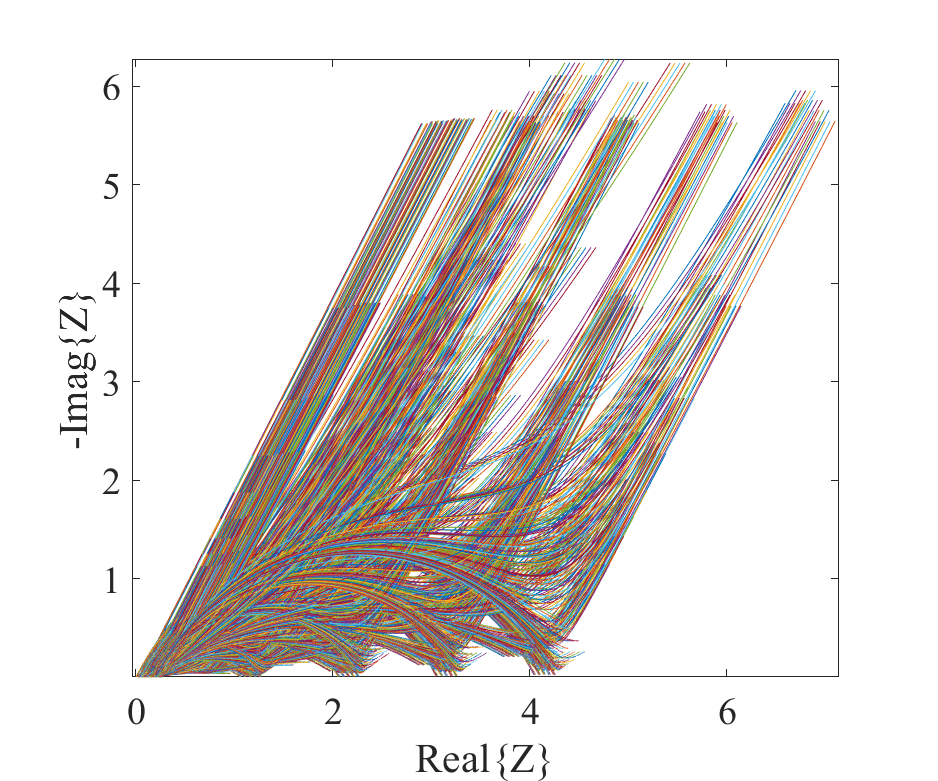}
\caption{The proposed method confirms that the FO ECM with two CPEs and Gr\"{u}nwald-Letnikov differentiation is globally structurally identifiable for batteries with the shown impedance spectra.}
\label{fig:eis}
\end{figure}

Structural identifiability of the FO-ECM in Fig.  \ref{fig:ecm2cpe} is then tested for various parameter sets chosen in the range of $R_\infty \in (0.01,0.2)$, $R_1 \in (0.05, 5)$, $C_1 \in (1,20)$, $C_2 \in (100,500)$, $\alpha_1 \in (0.1,0.9)$, $\alpha_2 \in (0.1,0.9)$. These ranges are commonly used in battery literature, \cite{Jacob2018}. It is observed that in all cases, there is a one-to-one map, which demonstrates the global identifiability of the EIS models in these ranges. Fig. \ref{fig:eis} shows the impedance spectra of undertaken models.

\section{Conclusions and future works}
In this paper, an efficient method was proposed for numerical structural identifiability analysis of fractional-order equivalent circuit models (FO-ECMs) with two constant phase elements (CPEs) under the Gr\"{u}nwald-Letnikov  differentiation, obtained through impedance spectroscopy. The proposed method confirms that the two-CPE FO-ECM in Fig. \ref{fig:ecm2cpe} is globally identifiable for a wide range of parameter values that is common in battery applications. Structural identifiability analysis for higher-order FO-ECMs and/or by using other approximations of fractional derivatives, as well as practical identifiability analysis, i.e., selection of the excitation signal, the effect of measurement noise, etc., are suggested for future research.

\ifCLASSOPTIONcaptionsoff
  \newpage
\fi

%

%
%
%




\end{document}